%% file: FDeMarchi_VERITAS_paper_2_submitted.tex
\documentclass[aps,prd,reprint,superscriptaddress,onecolumn]{revtex4-1}
\include{packages}

\begin{document}

\title{Testing the gravitational redshift with an inner Solar System probe:\\ the VERITAS case  } 
\author{Fabrizio De Marchi}
\email[Corresponding author: ]{fabrizio.demarchi@uniroma1.it}
\affiliation{Department of Mechanical and Aerospace Engineering, Sapienza University of Rome,   Via Eudossiana, 18, 00184 Rome, Italy}
\author{Gael Cascioli}
\affiliation{Department of Mechanical and Aerospace Engineering, Sapienza University of Rome,   Via Eudossiana, 18, 00184 Rome, Italy}
\affiliation{Currently at University of Maryland Baltimore County and NASA Goddard Space Flight Center, Maryland, USA}

\author{Todd Ely}
\affiliation{Jet Propulsion Laboratory, California Institute of Technology}
\author{Luciano Iess}
\affiliation{Department of Mechanical and Aerospace Engineering, Sapienza University of Rome,   Via Eudossiana, 18, 00184 Rome, Italy}
\author{Eric A. Burt}
\affiliation{Jet Propulsion Laboratory, California Institute of Technology}
\author{Scott Hensley}
\affiliation{Jet Propulsion Laboratory, California Institute of Technology}
\author{Erwan Mazarico}
\affiliation{NASA Goddard Space Flight Center, Maryland, USA}

\begin{abstract}
The NASA Discovery-class mission VERITAS, selected in June 2021, will be launched towards Venus after 2027. In addition to the science instrumentation that will build global foundational geophysical datasets, VERITAS proposed to conduct a technology demonstration for the Deep Space Atomic Clock (DSAC-2). A first DSAC successfully operated in low-Earth orbit for more than two years, demonstrated the trapped ion atomic clock technology, and established a new level of performance for clocks in space. DSAC-2 would have further improvements in size, power, and performance. It would host a $1\times{10}^{-13}$ grade USO to produce a frequency output with short-term stability of less than $2\times{10}^{-13}/\sqrt\tau$ (where $\tau$ is the averaging time).\\
However, due to funding shortfalls, DSAC-2, had to be canceled. The initially foreseen presence of an atomic clock on board the probe, however, raised the question whether this kind of instrumentation could be useful not only for navigation and time transfer but also for fundamental physics tests.
In this work, we consider the DSAC-2 atomic clock and VERITAS mission as a specific example to measure possible discrepancies in the redshift predicted by General Relativity by using an atomic clock onboard an interplanetary spacecraft. \\
In particular we investigate the possibility of measuring possible violations of the Local Lorentz Invariance and Local Position Invariance principles.\\
We perform accurate simulations of the experiment during the VERITAS cruise phase. We consider different parametrizations of the possible violations of the General Relativity, different operational conditions, and several different assumptions on the expected measurement performance. We show that DSAC-2 onboard VERITAS would provide new and improved constraints with respect to the current knowledge. \\
Our analysis shows the scientific value of atomic clocks like DSAC-2 hosted onboard interplanetary spacecraft.

\end{abstract}


\maketitle


\section{Introduction}\label{intro}
An important goal of modern physics is the unification of the four fundamental interactions under a theory that will encompass General Relativity (GR) and Quantum Mechanics. An overlying theory would converge to GR at large distances, where gravity is the dominant interaction, and to quantum theory at atomic/subatomic scales. Since the two theories are incompatible with each other (e.g., GR is a deterministic theory), a violation of GR is expected at a certain level. The search for possible violations of GR involves checks of the Equivalence Principle and, in a weak-field regime, measurements of the parameters of the parametrized post-Newtonian (PPN) 
 \cite{Will2014}. A deviation from the values predicted by GR for one or more of these parameters would indicate that GR is not the ultimate theory of gravitation.

The validity of GR has been challenged several times by astronomical measurements on an interstellar scale \cite{archibald2018} and in the solar system by radio tracking to interplanetary probes (e.g., \cite{bertotti2003a,genova2018,bernus2022}; see \cite{demarchi2020} for a summary).\\
Deviations from GR would appear as small deviations of the trajectories of the celestial bodies (or interplanetary probes) from the predictions. At the interstellar scale (e.g., stellar systems composed of compact objects) relativistic effects are strong and possible deviations from GR would be easier to detect. However, results are limited by uncertainties on masses and mutual distances of these objects.  On the contrary, in the solar system, despite very small relativistic effects (weak-field regime), the masses and trajectories of the principal bodies are known with high accuracy.\\
PPN parameters are measured through the solution of orbit determination problems based on ephemerides and on precise radiometric data (range and/or Doppler data) collected by interplanetary missions. Until now, no violations of GR have been found.\\
Range and range-rate data that will be acquired by VERITAS in its orbital phase may be used to perform tests of GR. The expected performances will, however, not be at the level of BepiColombo \cite{demarchi2020} because VERITAS is not endowed with a complete cancellation system for plasma noise (VERITAS enables only two simultaneous links - X uplink/X downlink and Ka uplink/Ka downlink, while BepiColombo has an additional X uplink/Ka downlink that enables full plasma noise cancellation).\\
However, we will show that a time transfer experiment during the cruise phase of VERITAS between the proposed onboard deep space atomic clock, DSAC-2 \cite{burt2021}, and the atomic clocks at the ground stations can provide precise measurements of both the gravitational redshift (i.e., the decrease/increase of the frequency of an electromagnetic wave as it exits/enters from/into a gravitational well) and of the Doppler shift due to the relative velocity between the emitter and the receiver.\\
This experiment has deep significance as it belongs to a class of tests involving the Equivalence Principle (EP), one of the two founding principles of GR. The fact that gravitation is a manifestation of the geometry of spacetime and not a “real” force, as the other three fundamental interactions, is an outcome of the universality of free fall, which is in turn a direct consequence of the EP.
The Einstein Equivalence Principle (EEP) is usually divided into three sub-principles:
\begin{itemize}
    \item 	Universality of Free Fall (UFF), which states that a test particle (i.e., an object with a negligible self-gravity) follows a trajectory in an external gravitational field independent of its mass or structure (i.e., trajectories depend only on the initial state). This sub-principle is often called the Weak Equivalence Principle (WEP).
    \item Local Lorentz Invariance (LLI), which states that the result of a local and non-gravitational experiment is independent of the orientation or velocity of the free-falling laboratory. 
    \item Local Position Invariance (LPI), which states that a non-gravitational experiment is independent of where it is performed in space and in time. 
\end{itemize}

By “non-gravitational experiments” we mean that in the laboratory the gravitational effects are negligible. Otherwise, the principle is called the Strong Equivalence Principle (SEP).\\
In other words, when comparing the free fall of two objects in the same laboratory (i.e., a “null test”), preferred-location and/or preferred-frame effects occur if the outcome depends on the location of the laboratory in an external gravitational field and/or its velocity in an inertial frame. If this occurs, the EEP is violated.\\
Conversely, if the EEP is valid, then the Doppler effect between two static clocks (or “gravitational redshift”) in the weak-field approximation is “universal,” independent of the nature of the clocks, and always given by $\Delta U/c^2$ \cite{turneaure1983} where $\Delta U$ is the difference between the Newtonian potentials at the clock locations and c is the speed of light.\\
While tests of the UFF/WEP are performed by Eötvös-like experiments, LLI and LPI are usually verified by comparing the frequency of two clocks in motion at different locations (i.e., gravitational potential) or times. The two clocks can be different (e.g., different materials and/or based on different physical processes to generate time) or identical. In this latter case, the experiment automatically excludes the coupling between gravitation and the interactions involved in clock function. \\
The sub-principles are expected to be connected: the so-called Schiff’s conjecture states that the validity of WEP guarantees the validity of LLI and LPI \cite{schiff1960}.\\
The violation of the EP can be parametrized by assuming that the inertia/energy of a particle is position- and velocity-dependent. Given a particle with rest mass $m_0$, position $\mathbf X$, velocity $\mathbf V$ in a gravitational potential $U$, the Lagrangian expanded to first order in $U$ and $V^2$ is given by \cite{wolf2016}
\be
L=-m_0c^2+\frac{1}{2}m_0\left(1+\frac{\delta m_I}{m_0}\right)V^2-m_0\left(1+\frac{\delta m_G}{m_0}\right)U\left(\mathbf{X}\right)
\ee
where $\delta m_G$, $\delta m_I$ represent the coupling of the (passive) gravitational and inertial masses/energies to the position and velocity of the frame, respectively. Using the Euler-Lagrange equations, the equations of motion are 
\be
\mathbf{a}=\mathbf{g}\left(1+\frac{\delta g}{g}\right)  
\ee{
}where 
\be 
\mathbf{g}=-\frac{\partial U}{\partial \mathbf{X}}          
\ee{
}and
\be
\frac{\delta g}{g}=\frac{\delta m_G}{m_0}-\frac{\delta m_I}{m_0}
\label{eq:4}
\ee
is the Eötvös parameter that can be measured in a dedicated experiment (e.g., the MICROSCOPE mission  \cite{touboul2017}).\\
Nordtvedt, 1975 (\cite{nordtvedt1975}) showed that the conservation of energy, in the presence of a WEP violation, would imply a small supplementary redshift in photons exchanged between two static identical clocks.  The gravitational redshift between identical receiver “r” and emitter “e” in case of a WEP violation being given by (weak-field approximation)
\be
\frac{\nu_r-\nu_e}{\nu_e}=\left(1+\alpha\right)\frac{U_e-U_r}{c^2}       
\ee
where $U_e$ and $U_r$ are the gravitational potentials at the emitter and the receiver, $\alpha=\left(\delta g/g \right)\left(m_0/\delta m\right)$, and $\nu_r$, $\nu_e$ are the receiver and emitter frequencies, respectively. 
The factor $m_0 / \delta m$  is the ratio of the atom rest mass-energy to its variation due to the specific quantum process involved in the generation of photons. In practice, for a given atomic clock, the factor $m_0 / \delta m$ depends on the type of atoms/molecules and on the type of transitions used to measure time. It is an “amplification factor” \citep{nobili2013} of the LPI violation parameters derived from the WEP.\\
In principle, different clocks have different “amplification factors” as the coupling with gravitation could be different. An analog factor is present in the second-order Doppler shift leading to the general formula for identical clocks
\be
\frac{\nu_r-\nu_e}{\nu_e}=\left(1+\varepsilon\right)  \frac{V_r^2-V_e^2}{{2c}^2}+\left(1+\alpha\right)\frac{U_e-U_r}{c^2}              
\ee
where $\varepsilon$ is the LLI violation parameter and $V_r$, $V_e$ are the velocities of the receiver and of the emitter with respect to an inertial frame. The validity of LLI and LPI entails $\varepsilon=\alpha=0$.\\
Due to the amplification factors, in the case of WEP violations, test masses involved in gravitational redshift and/or Doppler shift experiments (i.e., clocks) would be more affected than those used in Eötvös-type null-tests. \\
Following \cite{nordtvedt1975} we can obtain the LPI parameter after a direct Eötvös-type measurement. The amplification factor to be used depends on the model used. For example, if we assume that the WEP violation depends on electromagnetic interaction, the amplification factor is $\approx 10^3$ and, since the WEP violation parameter $\delta g / g$ is currently known at a $10^{-15}$ level  \citep{touboul2017}, we would obtain an accuracy on LPI parameter of about $10^{-12}$, a much more stringent limit than the current redshift tests accuracy.\\ 
On the contrary, if the coupling depends on the hyperfine transition, the direct measurement by clocks comparison would be the most accurate \cite{altschul2015}.\\
In 1976, Gravity Probe A (GP-A) was the first satellite to perform a redshift test in the gravitational field of the Earth by comparing two hydrogen masers. It confirmed the validity of LPI with an accuracy
$\sigma\left(\alpha\right)=1.4\times{10}^{-4}$ \cite{vessot1989}.\\
The satellites GSAT0201 and GSAT0202, were launched together in 2014 as part of the ESA Galileo constellation. Because of a technical problem, they were placed on an elliptic ($e$=0.162) orbit with a period of 13~h. Due to the eccentricity, this configuration was ideal for a gravitational redshift experiment. By comparing the two onboard passive hydrogen-masers, the experiment led to a factor 4.5 to 6 improvement ($\sigma(\alpha)=3.1 \times 10^{-5}$ \citep{herrmann2018} and $\sigma(\alpha)=2.49 \times 10^{-5}$ \citep{delva2018}) with respect to the GP-A results. 
A further improvement of about a factor 10 is expected from the upcoming Atomic Clock Ensemble in Space (ACES) mission \cite{savalle2019}.\\
The NASA Galileo interplanetary mission to Jupiter (launched in late 1989 with an USO onboard) performed the first redshift test based on the Sun’s gravitational field. The precision was $\sigma\left(\alpha\right)=5.0\times{10}^{-3}$ \cite{krisher1993}.\\
Regarding the LLI violation, the current uncertainty on $\varepsilon$ is at the $10^{-9}-10^{-8}$ level, obtained by ground laboratory experiments \cite{tobar2010,botermann2014}. \\
In this work we analyze the opportunity to perform at the same time a gravitational redshift (LPI violation) and Doppler shift (LLI violation) experiment, during the cruise phase of VERITAS. Our simulations show that the exquisite stability of an onboard technology demonstrator such as DSAC-2 would enable a factor 2.5 to 5 improvement in the uncertainty of   $\alpha$. Conversely, the LLI violation test cannot improve the current accuracy at the ${10}^{-8}$ level.\\
This paper is structured as follows: \sref{sec:2} describes the VERITAS mission and the atomic clock DSAC-2, and in \sref{sec:3} we outline the theoretical framework to be used in the experiment. In \sref{sec:4} we describe our simulations and we discuss the results. Finally, in \sref{sec:5} we draw the implications and conclusions of the outcomes of this work.

\section{Experiment concept}\label{sec:2}
\subsection{VERITAS mission}
VERITAS (for Venus Emissivity, Radio Science, InSAR, Topography, and Spectroscopy) is a NASA Discovery mission planned for launch after 2027. 
VERITAS would host a set of instruments designed to advance our understanding of Venus, namely a synthetic aperture radar (VISAR; \cite{hensley2020}), an infrared spectrometer (VEM; \cite{helbert2020}) and the Integrated Deep Space Transponder (IDST) enabling the gravity science investigation (\cite{cascioli2021,cascioli2022,mazarico2019,iess2021agu}). The nominal mission scenario consists of a 7-month cruise phase, followed by planetary capture in a highly eccentric orbit with period $\approx 13$~h. An aerobraking campaign then reduces the orbital period and eccentricity. This first aerobraking campaign is followed by an initial science phase in an elliptical orbit (science phase 1, SP1) mainly designed for VEM observations. SP1 is followed by another 5 months of aerobraking, used to circularize the orbit, and finally the main observation campaign covering 4 Venus sidereal rotation periods (or cycles, $\approx 243$ days each) in which VISAR, VEM, and gravity science measurements will be collected (science phase 2, SP2). \\
\begin{center}
\begin{figure}[h!]
\includegraphics[width=.4\columnwidth]{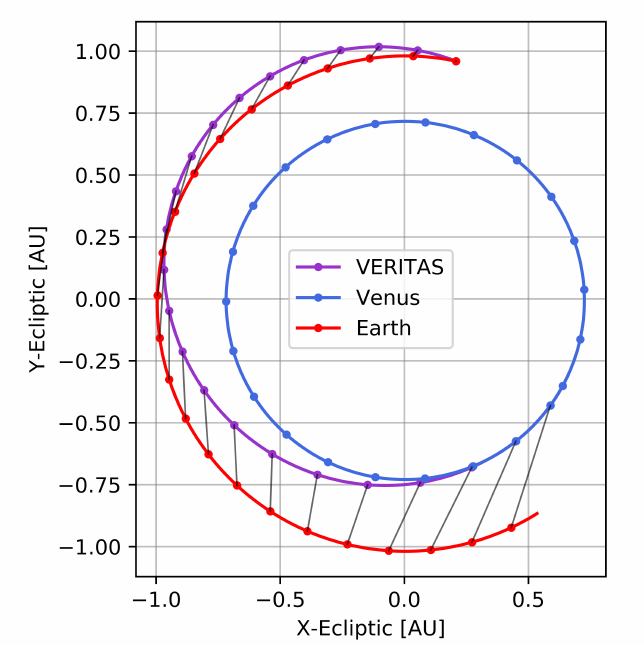}\\
\includegraphics[width=.4\columnwidth]{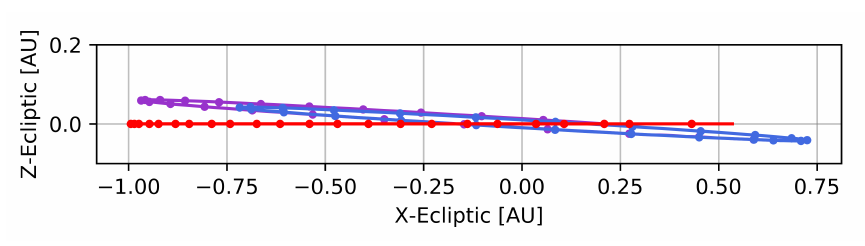}
\caption{\footnotesize VERITAS cruise trajectory in the ecliptic plane (top panel) and in the X-Z plane (lower panel). Markers indicate ten-day periods. The black segments show the Earth-VERITAS distance during the cruise.}
\label{fig:fig1}
\end{figure}
\end{center}
In this work we are 
interested in the cruise phase. \fref{fig:fig1} shows the path of the spacecraft during the transfer from Earth to Venus. The spacecraft will be inserted on a direct transfer orbit inclined on the ecliptic plane, to match the Venus $\sim 4$~deg. orbit inclination. During the cruise there will be no superior solar conjunctions (SSCs). This implies that the radiometric observables collected in the context of the redshift experiment would not be substantially affected by solar plasma noise. This fact, coupled with the planning of a very limited amount of orbit adjustment maneuvers, allows for a very quiet dynamical and observational environment of the probe, ideal for testing faint perturbations on the propagation of radio waves related to possible LLI/LPI violations.\\
The radiometric tracking of the probe will rely on the onboard tracking system, close heritage of ESA’s BepiColombo \cite{iess2021,iess2009,cascioli2021}. The VERITAS tracking system is able to establish simultaneous two-way tracking links with Earth stations in the X- and Ka-bands (8 and 32~GHz respectively). Operational tests of the tracking equipment onboard BepiColombo allowed verification of its excellent performance at Ka-bands, which are on the order of 0.02~mm/s at 10~s integration time for range rate and 1-2~cm for range (2~s sampling rate) \cite{cappuccio2020b}. During the VERITAS cruise, however, the Ka-band would not be baselined, as routine tracking operations will be conducted through single X-band links with reduced performance. Operational assessments during the BepiColombo cruise phase have shown an X-band accuracy on the order of 0.1~mm/s and 0.04~mm/s at 10~s and 60~s integration times, respectively (Paolo Cappuccio, private communication).

\subsection{DSAC-2}\label{sec:2b}
\begin{center}
\begin{figure}[h!]
\includegraphics[width=.6\columnwidth]{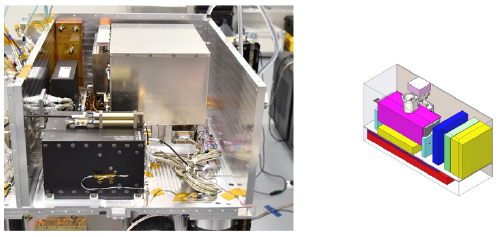}
\caption{\footnotesize On the left is a photo of the DSAC-1 instrument (silver box) during space craft integration (a GPS receiver, shown in the foreground, will not be part of DSAC-2). On the right the DSAC-2 system concept is shown at the same scale.}
\label{fig:fig2}
\end{figure}
\end{center}

NASA’s Deep Space Atomic Clock (DSAC) Technology Demonstration Mission recently completed its two-year mission in low-Earth orbit (operations ended on September 18, 2021). DSAC successfully showed the technology’s viability for sustained, reliable operations and for providing the most stable frequency ever demonstrated in space ($\approx 3\times 10^{-15}$ at one-day and a drift $<3 \times 10^{-16}$/day \cite{burt2021}). This success warranted development of a next-generation DSAC, called DSAC-2.\\
The DSAC-2 conceptual design intended to use less power, to be smaller, and to be longer-lived than DSAC-1, while maintaining excellent performance. With DSAC-2 operating as an external reference to VERITAS’ IDST, it would have been possible to collect X-band, Ka-band, and combined X/Ka one-way downlink Doppler to support this proposed GR test. This study examines the GR tests and outcomes that would have been possible if funding for the DSAC-2 TDO were maintained.\\
An important aspect of the DSAC-2 project is to advance the trapped ion clock technology beyond DSAC-1 to include: 1) superior stability, 2) increased lifetime, 3) reduced size, weight, and power (SWaP), and 4) improve the fabrication yield percentage and robustness. DSAC-2 would have a stability of  $2 \times 10^{-13}/\sqrt{\tau}$ in the short term and  $3 \times 10^{-15}$ at one day - similar to the DSAC-1 stability achieved on the ground. To address the DSAC-2 SWaP requirements of 13~L, 13~kg, and 42~W, DSAC-2 would simplify the ion trap and frequency chain architectures used in DSAC-1, thereby reducing components and size as well as power, with current design estimates falling well below these requirements. \fref{fig:fig2} shows a scale comparison of DSAC-1 and DSAC-2. In addition to design simplification, use of commercial off-the-shelf (COTS) parts and removal of high tolerances where possible will increase instrument yield and reliability.

\section{Mathematical framework }\label{sec:3}
\begin{center}
\begin{figure}[h!]
\includegraphics[width=.56\columnwidth]{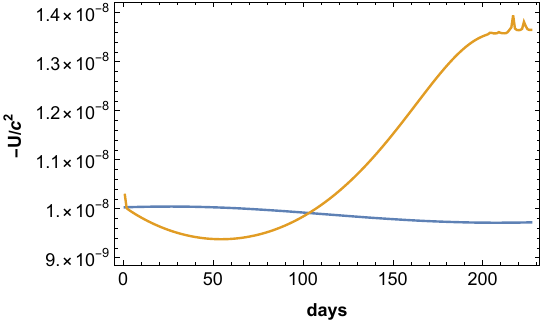}
\includegraphics[width=.43\columnwidth]{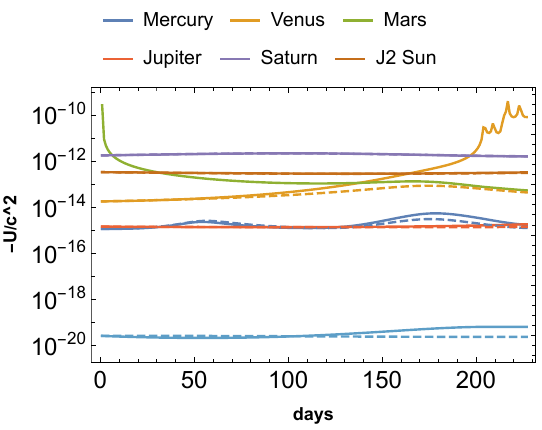}
\caption{\footnotesize LPI signals. Left panel: total (Sun+planetary) LPI contributions for the ground station (blue) and probe (yellow) clocks. Right panel: different contributions to the ground station (dashed) and probe (solid) clocks.}
\label{fig:signals}
\end{figure}
\end{center}

The relativistic Doppler effect is a combination of the classical Doppler effect and of time dilation. \\
Given a station “st” (the “receiver”) and a spacecraft “sc” (the “emitter”), the relativistic Doppler shift is \cite{bertotti2003b,wolf1995}
\be
\frac{\nu_{st}}{\nu_{sc}}=\left(\frac{1-\mathbf{n}\cdot\mathbf{v}_{st}/c}{1-\mathbf{n}\cdot\mathbf{v}_{sc}/c}\right)\ \ \frac{d\tau_{sc}}{d\tau_{st}}
\label{eq:7}
\ee
where $\mathbf{n}$ is the unit vector along the propagation path (i.e., from the spacecraft to the station)
\be
\mathbf{n}=\frac{\mathbf{r}_{st}-\mathbf{r}_{sc}}{\left|\mathbf{r}_{st}-\mathbf{r}_{sc}\right|}
\ee
and  $\mathbf{r}_{st},\mathbf{r}_{sc},\mathbf{v}_{st},\mathbf{v}_{sc}$ are their respective positions and velocities in an inertial frame.\\
The first term on the right-hand side in \eref{eq:7} represents the “classical” Doppler effect while the second is the time dilation factor (or “transverse Doppler effect”).\\
Referring to Barycentric Coordinate Time (TCB), the time coordinate for the Barycentric Celestial Reference System (BCRS), we estimate the time dilation factor as
\be
\frac{d\tau_{st}}{d\tau_{sc}}=\frac{d\tau_{sc}}{dt_{TCB}}/\frac{d\tau_{st}}{dt_{TCB}}
\label{eq:9}
\ee
$t_{TCB}$ represents the time of a clock moving as the Solar System barycenter but not influenced by the gravitational potential caused by the Solar System.\\
The Schwarzschild metric in isotropic coordinates and in the weak field approximation, reads
\be
ds^2=\left(c\ d\tau_{sc}\right)^2\approx\left(1+2U_{sc}\right)\left(c\ dt\right)^2-\left(dx^2+dy^2+dz^2\right)
\label{eq:10}
\ee
where $U_{sc}$ is the gravitational potential at the spacecraft. From Eqs. (\ref{eq:9}) and (\ref{eq:10}), we obtain
\be
\frac{d\tau_{sc}}{dt_{TCB}}=\sqrt{1+\frac{2U_{sc}-v_{sc}^2}{c^2}}
\ee
with $v_{sc}^2=\left|\mathbf{v}_{sc}\right|^2=\left(dx^2+dy^2+dz^2\right)/dt^2$.\\
Applying this method to both station and spacecraft, the time dilation reads
\be
\frac{d\tau_{sc}}{d\tau_{st}}=\sqrt{\frac{1+\left(2U_{sc}-v_{sc}^2\right)/c^2}{1+\left(2U_{st}-v_{st}^2\right)/c^2}}
\ee
where $U_{st}$ is the gravitational potential at the ground station.\\
The relativistic Doppler formula is therefore (in the weak-field approximation)
\be
\frac{\nu_{st}}{\nu_{sc}}=\sqrt{\frac{1+\left(2U_{sc}-v_{sc}^2\right)/c^2}{1+\left(2U_{st}-v_{st}^2\right)/c^2}}\left(\frac{1-\mathbf{n}\cdot\mathbf{v}_{st}/c}{1-\mathbf{n}\cdot\mathbf{v}_{sc}/c}\right).
\ee
Finally, the expansion up to $c^{-2}$ yields \cite{krisher1993} (\cite{krisher1993} assumes positively defined $U_{sc}$ and $U_{st}$, while we assume that gravitational potential terms are negative)
\be
\frac{\Delta\nu}{\nu_{sc}}=\frac{\nu_{st}-\nu_{sc}}{\nu_{sc}}\approx-\frac{\mathbf{n}\cdot\left(\mathbf{v}_{st}-\mathbf{v}_{sc}\right)}{c}-\frac{\left(\mathbf{n}\cdot\mathbf{v}_{sc}\right)\left(\mathbf{n}\cdot\mathbf{v}_{st}\right)}{c^2}+\frac{\left(\mathbf{n}\cdot\mathbf{v}_{st}\right)^2}{c^2}+\frac{v_{st}^2-v_{sc}^2}{2c^2}+\frac{U_{sc}-U_{st}}{c^2}.
\label{eq:14}
\ee
Two-way Doppler measurements do not sense (at the $c^{-2}$ level) the relativistic contributions (i.e., the last two terms in \eref{eq:14}) due to mutual cancellation. To detect the signals from the last two terms, two clocks at both ends of a one-way radio link are necessary.\\
In order to account for the most general case with different clocks, we adopt the following parametrization for the redshift signal due to LPI and LLI violations
\be
\frac{\Delta\nu}{\nu_{sc}}\left(\alpha_{sc},\alpha_{st},\varepsilon_{sc},\varepsilon_{st}\right)=\left(1+\alpha_{sc}\right)\frac{U_{sc}}{c^2}-\left(1+\alpha_{st}\right)\frac{U_{st}}{c^2}+\left(1+\varepsilon_{st}\right)\frac{v_{st}^2}{2c^2}-\left(1+\varepsilon_{sc}\right)\frac{v_{sc}^2}{2c^2}.
\label{eq:15}
\ee
The parameters $\alpha_{sc},\alpha_{st},\varepsilon_{sc},\varepsilon_{st}$ are by definition zero in GR.\\
In \fref{fig:signals} we report the Doppler signals of these parameters. The LPI signals $-U_{sc}/c^2$ and $-U_{st}/c^2$ have been calculated including solar and planetary contributions. Planetary contributions (between $10^{-15}$ and $10^{-10}$) as well as the Sun’s $J_2$ influence ($10^{-20}$)  are negligible with respect to the Sun's potential.
We will consider 4 cases: 1) LPI and LLI test with different clocks: the most general case 2) LPI test only based on different clocks. 3) LPI and LLI test based on identical clocks and 4) LPI test only and based on identical clocks (hereafter, “Delva-like” configuration).

\section{Simulations and results}\label{sec:4}
\begin{center}
\begin{figure}[h!]
\includegraphics[width=.4\columnwidth]{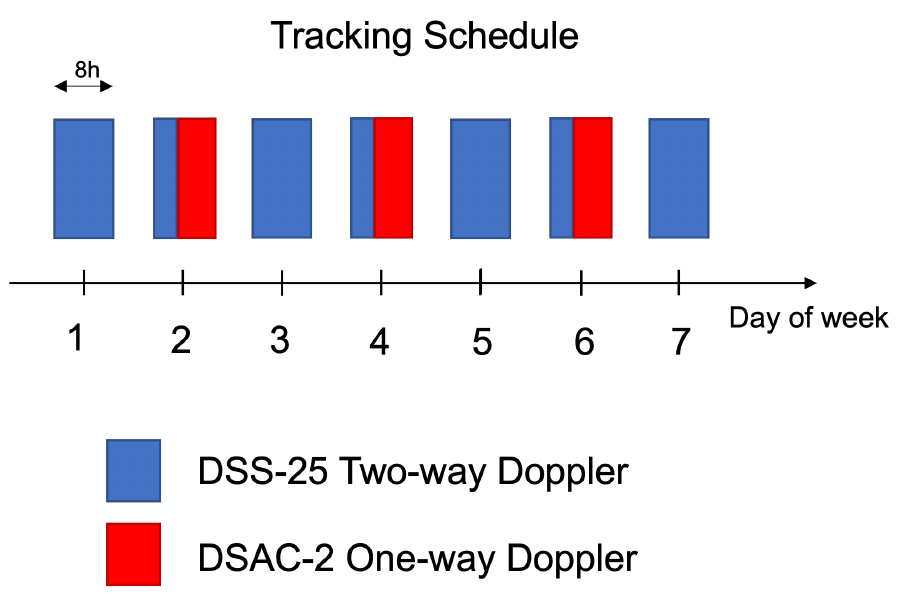}
\caption{\footnotesize Simulated tracking schedule.}
\label{fig:fig3}
\end{figure}
\end{center}

\begin{center}
\begin{figure}[h!]
\includegraphics[width=.8\columnwidth]{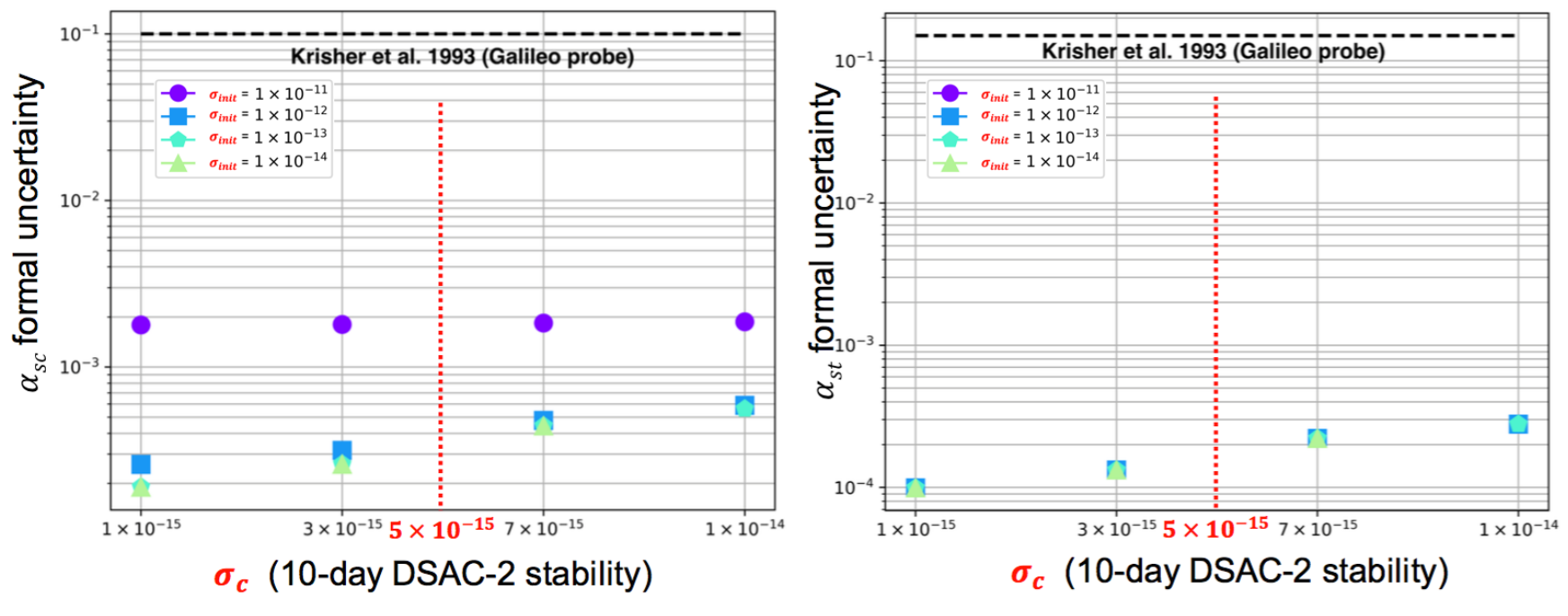}
\includegraphics[width=.8\columnwidth]{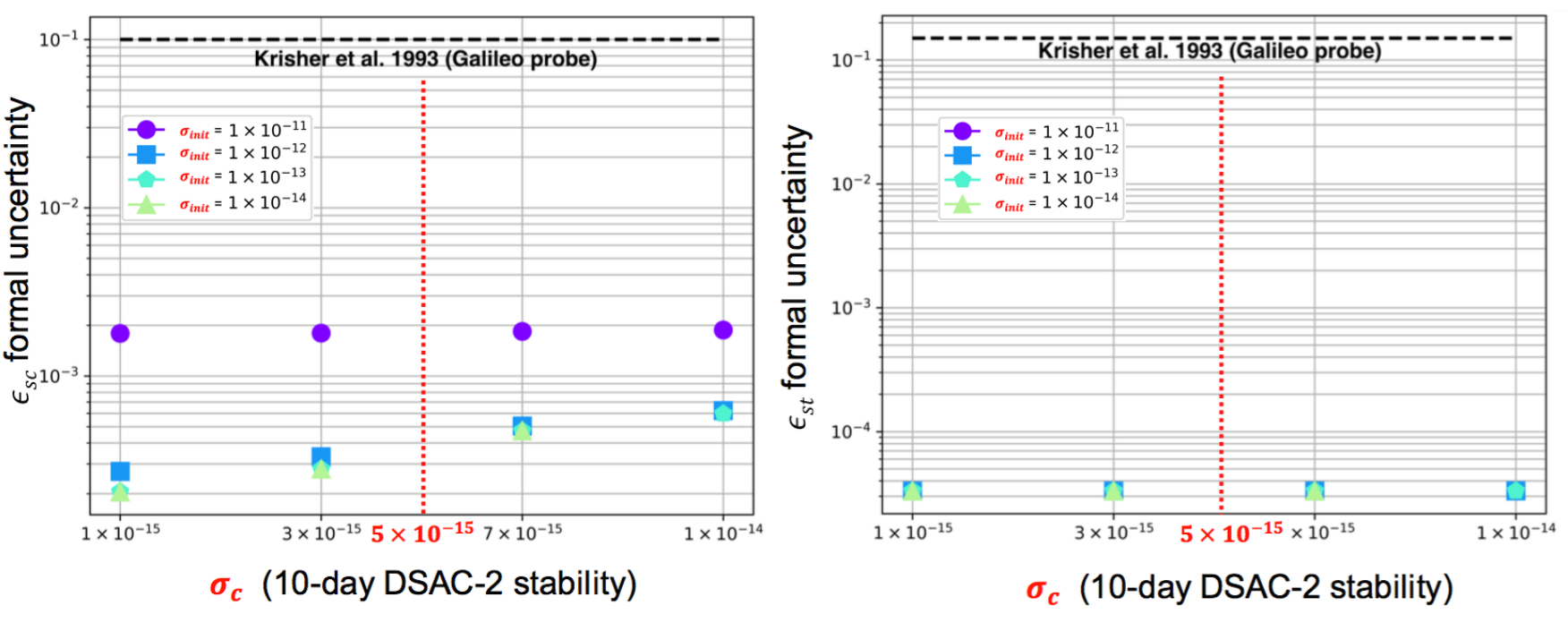}
\caption{\footnotesize Case I. Simulation results in terms of formal uncertainty on the parameter of interest (y-axis), initial frequency bias uncertainty $\sigma_{init}$ (x-axis) with varying assumptions for the 10-day frequency stability of DSAC-2, $\sigma_c$ (colored markers). Top (bottom): expected formal uncertainties for LPI (LLI) violation parameters.  All results are compared with the Galileo probe experiments outputs. The dotted vertical line represents the conservative upper limit ($5\times{10}^{-15}$) for the $\sigma_c$ constraint. The dotted horizontal lines report the results by \cite{krisher1993}.}
\label{fig:fig4}
\end{figure}
\end{center}

\begin{center}
\begin{figure}[h!]
\includegraphics[width=.9\columnwidth]{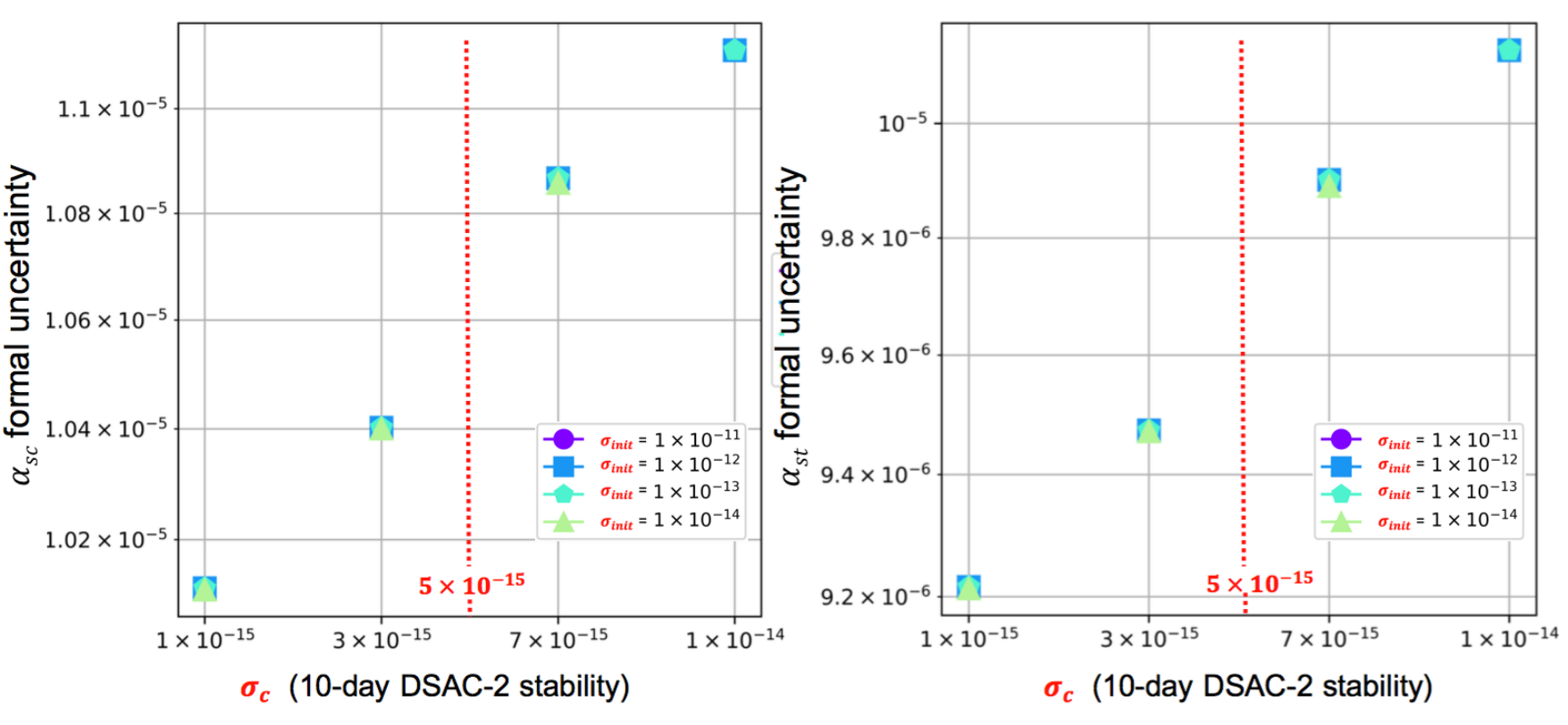}
\caption{\footnotesize Case II. LPI violation test with different clocks. Left (right): expected formal uncertainty for the spacecraft (ground station) clocks.}
\label{fig:fig5}
\end{figure}
\end{center}

\begin{center}
\begin{figure}[h!]
\includegraphics[width=.85\columnwidth]{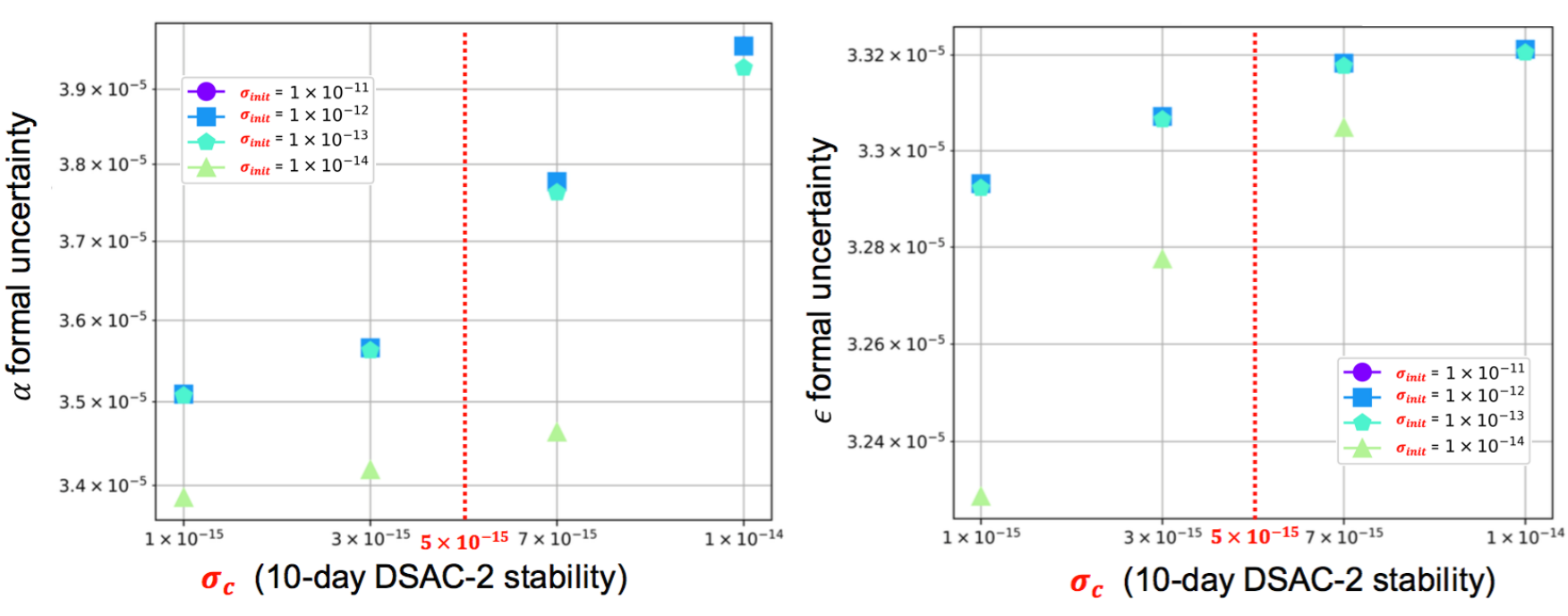}
\caption{\footnotesize Case III. LPI and LLI violation test with identical clocks. Left (right): expected formal uncertainty for LPI (LLI) violation parameter.}
\label{fig:fig6}
\end{figure}
\end{center}

\begin{center}
\begin{figure}[h!]
\includegraphics[width=.45\columnwidth]{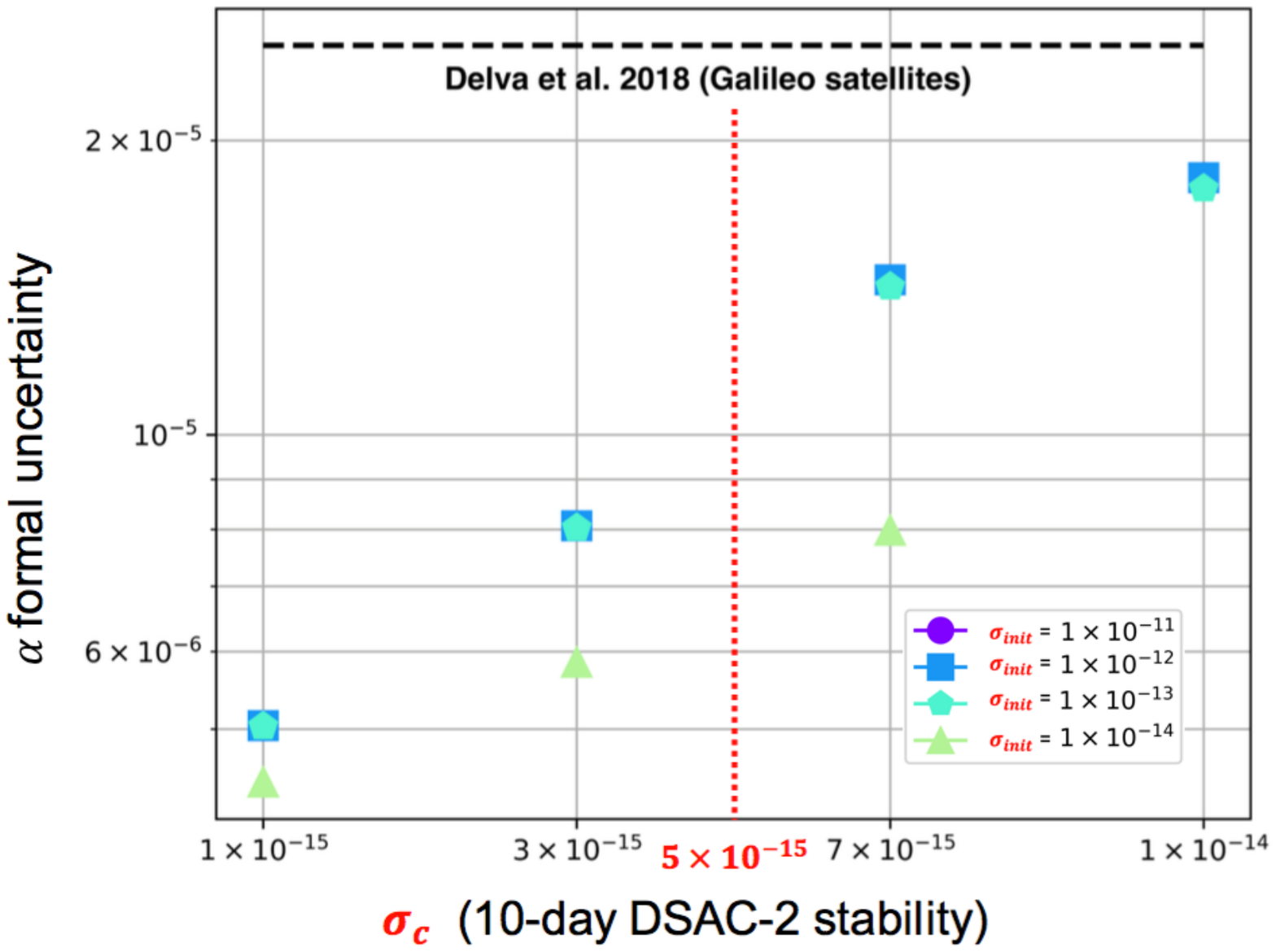}
\caption{\footnotesize Case IV. Simulation of the LPI violation test with identical clocks. The dotted horizontal line reports the current knowledge of the $\alpha$ parameter by \cite{delva2018}.}
\label{fig:fig7}
\end{figure}
\end{center}

We conducted an extensive set of numerical simulations to assess the expected performance of the proposed experiment. The simulations have been conducted using the NASA-JPL MONTE orbit determination software library \cite{evans2018}. We assume spacecraft tracking from Earth for 8 hours/day - 7 days/week.  Four days a week are reserved for standard 8 hours/day DSN two-way X-band tracking, while three days a week for 3 hours of DSN two-way tracking followed by 5 hours of DSAC-2 one-way downlink tracking (see \fref{fig:fig3}).  We assume the Doppler tracking to be supported by the station DSS-25 of the NASA DSN located in Goldstone (CA). For the two-way X-band tracking data noise we assume an RMS of 0.04~mm/s at 60~s integration time in accordance with operational observations on BepiColombo (see \sref{sec:2}). Note that Ka-band tracking would reduce the Doppler noise to 0.01-0.02~mms/s at the same integration time. An estimate of the noise level of the DSAC-2 observables (one-way downlink Doppler) has been obtained by performing a detailed noise budget, using the same assumptions for noise sources (other than the intrinsic stability of the DSAC-2 clock) used for the VERITAS error budget \cite{smrekar2022}.\\
For the intrinsic stability of the DSAC-2 frequency standard we assumed a white noise scaling on short integration timescales (under a day), closely in accord with measured performances of DSAC. The expected noise characteristics for timescale from about 30 seconds to a day are
\be
\sigma_y=\frac{2\times{10}^{-13}}{\sqrt\tau}
\ee
where $\sigma_y$ is the Allan standard deviation (a figure of merit for the fractional frequency stability $y=\Delta f/f)$ and $\tau$ is the averaging time. \\
At long averaging times, the expected clock noise floor will be about  ${10}^{-15}$ most likely due to long-term variations in the environment.\\
Given the verified DSAC-1 performances and planned enhancements, DSAC-2 is expected to be extremely insensitive to environmental perturbations. In particular:
\begin{itemize}
    \item The fractional frequency temperature sensitivity of DSAC-1 was about $10^{-14}/^\circ$C with no thermal regulation (this result is unique among space clocks). The planned enhancement of DSAC-2 would push this limit to  $2\times 10^{-15}/^\circ$C. Moreover, in interplanetary space the temperature variations are expected to be smaller than in an orbital environment (probably by an order of magnitude or more).
    \item 	The magnetic sensitivity of DSAC-1 was about $10^{-14}/$G but, again, magnetic variations during interplanetary cruise should be almost absent.
    \item No frequency shift effects due to radiation were observed in DSAC-1 which flew in a radiation environment significantly worse (South Atlantic Anomaly) than would be experienced in deep space.  The exception is during solar events, but it is likely that data during such events would be discarded.
\end{itemize}
For these reasons, the cruise phase of an interplanetary probe represents an ideal environment to perform the experiment here proposed.\\
In our simulations we assume an integration time of 60 seconds for both the DSAC-2 one-way Doppler and the two-way Doppler observables { (Ka and X band, respectively)}.\\ 
For the DSAC-2 observables, we considered the Doppler error budget estimation reported in \cite{smrekar2022} that takes into account several sources (e.g., plasma, troposphere and ground system components). In particular, plasma noise is expected to be relatively mild because of the small Earth-spacecraft distance in cruise and no superior conjunctions are expected. The tropospheric contribution, given the level of performance demonstrated at DSS-25 during Cassini data analysis using the DSN media calibration system (MCS) \cite{keihm2004}, is expected to be $<20$\% of the total. The expected noise level at 60~s is on the order of $\sigma_y\approx 3 \times 10^{-14}$.
The numerical simulations are conducted by simulating the tracking data and the DSAC-2 data based on the current design trajectory accounting for the above-mentioned assumptions, computing the partial derivatives of the observations with respect to the parameters of interest and by inverting the classical orbit determination normal equations (see e.g., \cite{tapley2004}).\\
We simulate the experiment during the 15-week period free of maneuvers to minimize the number of dynamical discontinuities in the trajectory. We use the ORACLE orbit determination filter developed at Sapienza University which has been successfully validated and applied to many data analyses and simulations (e.g., \cite{iess2018,durante2020,cascioli2020}). The dynamical model employed to simulate the trajectory of the probe accounts for the gravitational interaction of the main solar system bodies and the radiation pressure imparted by the solar photons on the probe. To simulate the real data analysis approach, we have run two distinct sets of simulations employing two different orbit determination frameworks. We have conducted a multi-arc analysis and a single arc analysis. In the multi-arc analysis, we divide the trajectory into 10-days arcs and treat each period as a separate data arc. This method entails two separate sets of solve-for parameters: {\em local} parameters and {\em global} parameters. The local parameters are the ones affecting only one arc, while the global ones are common to all the arcs. \\
The set of estimated local parameters include the state of the probe (position and velocity at the beginning of the arc) a solar radiation pressure (SRP) scale factor. The global parameters include the gravitational redshift-related parameters ($\alpha,\epsilon$) and the position and velocity of the Earth with a large, unconstraining, a priori uncertainty of 50~km and $10^{-8}$~km/s, respectively, derived from \cite{demarchi2020}. These assumptions for the a priori uncertainties account for possible errors in Earth’s ephemerides that would lead to systematic effects. For the same reason, we also assumed a large a priori uncertainty (25\%) for the SRP scale factor.\\
To introduce the methodology used to account for the possible fluctuations in the frequency stability of the DSAC-2, a discussion on its behavior during the experiment is required. At the beginning of the experiment DSAC-2 will be switched on; the reference frequency of the DSAC-2, after a calibration phase, will be known with a certain precision and accuracy, depending on the internal calibration hardware. Indeed, the actual DSAC reference frequency ($f_0$) will differ from the expected one ($f_0^e$) by a frequency bias ($B_0$) such that
\be
f_0=f_0^e+B_0.
\ee
After the calibration phase, DSAC-2 will continue running with a stability as discussed in \sref{sec:2b}. Indeed, at the beginning of the i-th arc we can expect the reference frequency to be into the interval
\be
f_{i-1}-f_{i-1}\sigma_y<f_i<f_{i-1}+f_{i-1}\sigma_y .
\ee
Where $f_{i-1}$ is the reference frequency of the previous arc and $\sigma_y$ is the fractional frequency stability (in this case corresponding to $\tau$=10 days).
In terms of biases:
\be
B_{i-1}-f_{i-1}\sigma_y<B_i<B_{i-1}+f_{i-1}\sigma_y .
\label{eq:19}
\ee
In the orbit determination filter, then, we solve for the initial bias with an a priori uncertainty $\sigma_{init}$ in line with the predicted calibration capability and, in each arc, we solve for the local frequency bias. The biases of each arc, however, are not unrelated, thus in each arc we estimate the local frequency bias subject to the inter-arc constraint of \eref{eq:19} with an a priori uncertainty $(\sigma_c)$ in line with the predicted 10-day frequency stability (i.e., the inter-arc constraint reads $\left|\left|B_i-B_{i-1}\right|\right|<\sigma_cf_{i-1})$. \\
The single-arc analysis reproduces the multi-arc frequency bias management approach by modeling the frequency bias as a stochastic process. We model the frequency bias as a white noise uncorrelated stochastic process with an update time of 10 days. The a priori uncertainty on the initial value of the stochastic parameter is set to $\sigma_{init}$ and the stochastic process uncertainty is set to $\sigma_c$.\\
The simulations presented below are performed to explore different assumptions regarding the Doppler shift parametrization (accordingly to \eref{eq:15}) while estimating only $\alpha$ or both $\alpha,\varepsilon$ together, with different assumptions regarding the type of clock equipped at the DSN receiver (i.e., an atomic clock based on the same atom and on the same internal atomic transition as the DSAC). We also explore the redshift-related parameter retrieval varying the stability assumptions on DSAC ($\sigma_{init}$ and $\sigma_c$).\\
The range of selected values of $\sigma_{init}$ is based on the DSAC-2 retrace value and characterized on the ground.  It is expected to be between ${10}^{-14}$ and ${10}^{-13}$ (this latter is a very conservative assumption). The simulations presented here were performed on an even broader spectrum, ranging from ${10}^{-14}$ to ${10}^{-11}$. \\
The parameter $\sigma_c$ corresponds to the 10-day ADEV value, an integration time where the clock is expected to be largely limited by its noise floor.  As explained above, enhancements planned for DSAC-2, as well as the relatively benign environment in space, should bring this below the  $10^{-15}$ level.\\ 
The DSAC-1 flight experiment achieved a $\sigma_c < 3 \times 10^{-15}$. A drift of $3\times 10^{-16}$/day was estimated by linear fit to the DSAC-1 frequency residuals. This fit includes flicker and random walk noise that, at 1 day timescale, can appear similar to  drift.  For this reason, a $3 \times 10^{-15}$/10~days noise level represents a conservative upper bound. Moreover, to account for even more pessimistic scenarios (and to reduce also eventual systematic effects), for the simulations we assumed $10^{-15}<\sigma_c<10^{-14}$.  The two approaches (single/multi-arc orbit determination) led to substantially identical results. 
The figures reported are derived from the stochastic approach.\\
The results by \cite{krisher1993} reflect the capabilities enabled by the previous generation of tracking systems. We explore the expected uncertainty retrieved by VERITAS under four different parametrization assumptions to provide a wide array of possible outcomes based on the real data analysis strategy employed.\\
The first case, Case I, we considered is the most general one: LPI and LLI violation with different clocks (4 parameters to be estimated). The results are reported in \fref{fig:fig4}.\\
In the second case, Case II, we considered the LPI violation only, tested with different clocks (\fref{fig:fig5}). The uncertainties are about 2.5 smaller than the current uncertainty \cite{delva2018}.\\
As a third case, Case III, we considered a parametrization employing both $\alpha$ and $\varepsilon$ but assuming that atoms and atomic transitions employed in the ground receiving clock are the same as that used by DSAC. In this case we do not show a “state-of-the-art line” as, to our knowledge, no previous experiment has met these assumptions: the expected uncertainties are about ${3\ \times10}^{-5}$ in both cases (and shown in \fref{fig:fig6}). Regarding the LLI, this result is about 3 orders of magnitude worst that the current uncertainty \cite{tobar2010,botermann2014}.\\
The last case, Case IV,  explores a Delva-like configuration (see \sref{sec:3}) in which a single $\alpha$ is estimated (i.e., a test of the LPI violation with identical clocks). Results are plotted in \fref{fig:fig7}. The expected formal uncertainty of the parameter $\alpha$, compared with the results of \cite{delva2018}, shows an improvement of a factor 2.5-5, a quite significant amelioration.\\
This can be ascribed to a set of reasons: the Galileo satellite experiment lasted 3 years while the measurement described here lasts approximately 200 days, but interplanetary LPI signals are about 20 times larger (as shown in \fref{fig:signals}). Furthermore, the enhanced clock stability of DSAC-2 (e.g., relative to Galileo) enables the long measurement times needed to realize the experiment over large distances. Finally, as described above, DSAC-1 had a noise floor of about $3\times 10^{-15}$ and a drift of $3\times 10^{-16}$/day. Galileo clocks (before drift-removal) drift at about $10^{-14}$/day and have a higher noise floor.\\
The LPI and LLI parameters are in general strongly correlated. In \aref{sect:appendix} we report the correlation plots (\fref{fig:corr}) in the four cases considered and assuming $\sigma_{init}=10^{-14}$ and $\sigma_c=3\times 10^{-15}$. Different assumptions lead to near identical plots.


\clearpage
\section{Conclusions}\label{sec:5}
In this work we investigated the possibility to perform an advanced test of Local Lorentz and Local Position invariance by an interplanetary spacecraft with a precise atomic clock onboard. As a specific example, we simulated the experiment where the Doppler signal is generated by a DSAC-2 onboard the VERITAS probe during the cruise phase to Venus.\\
The scarcity of planned maneuvers and gravity assists make this phase particularly suited for this kind of measurement. Moreover, the atomic clock DSAC-1, placed in a 100 minute Earth orbit, demonstrated a high insensitivity to thermal, magnetic and radiation effects \cite{burt2021}. These perturbations are expected to be much smaller in the interplanetary environment.\\
Our simulations show that these gravitational redshift and transverse Doppler (time dilation) experiments would benefit from the excellent stability of an onboard DSAC-2 atomic clock and would serve as a test of General Relativity predictions.\\
In our simulations we considered the possibility to test both LLI and LPI violation. The LLI is currently verified to a $10^{-9}-10^{-8}$-level \cite{tobar2010,botermann2014}. An experiment with VERITAS would not provide any improvement since the expected accuracy is about 3 orders of magnitude bigger. For the LPI, our simulations show that the experiment with VERITAS could improve the current accuracy on LPI by a factor between 2.5 and 5, depending on the assumptions on the DSAC-2 stability.\\
To conclude, we recognize that such an experimental setup could be well suited for future tests of LPI in the Solar System and recommend the consideration of hosting atomic clocks (with performances similar or better than DSAC-2) in future interplanetary probe designs.

\begin{acknowledgments}
FDM, GC and LI acknowledge support from the Italian Space Agency under contracts 2020-15-HH.0 and 2022-15-HH.0. 
A portion of the work described in this paper was carried out at the Jet Propulsion Laboratory, California Institute of Technology, under a contract with the National Aeronautics and Space Administration (80NM0018D0004).
The authors are thankful to S. Smrekar for useful discussions.
FDM would like to thank N. Bartel for the fruitful discussion at the COSPAR 2022 Assembly.\\
The authors finally thank the anonymous referee for their valuable review, which provided improvements and suggestions for this paper and future work. 
\end{acknowledgments}

\appendix
\section{}\label{sect:appendix}
We report here the correlation plots for the four cases described. The high correlations among LPI and LLI parameters represent a limit for the most general cases. On the contrary, LPI and LLI parameters are scarcely correlated with the other global parameters.
\begin{center}
\begin{figure}[h!]
\includegraphics[width=.4\columnwidth]{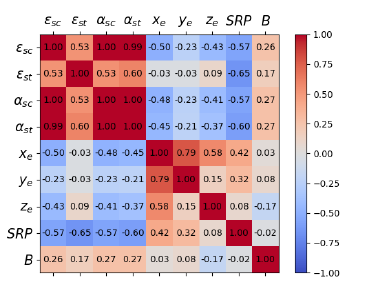}
\includegraphics[width=.4\columnwidth]{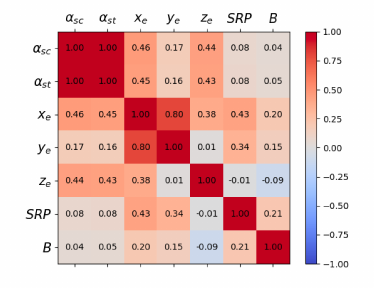}
\includegraphics[width=.4\columnwidth]{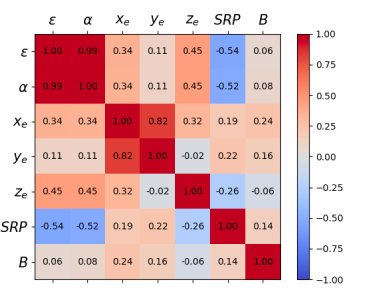}
\includegraphics[width=.4\columnwidth]{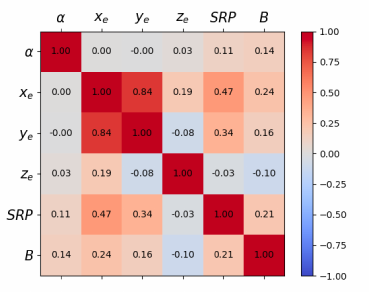}
\caption{\footnotesize Correlation plots for the covariance analysis described in Sect. IV: case I (top left), case II (top right), case III (bottom left) and case IV (bottom right). SRP is the solar radiation pressure scale factor, B is the stochastic frequency bias and ($x_e, y_e, z_e$) are the coordinates of the Earth. Assumptions: $\sigma_{init}=10^{-14}$ and $\sigma_c=3\times1 0^{-15}$.}
\label{fig:corr}
\end{figure}
\end{center}

\clearpage

\end{document}

%% file: packages.tex
\usepackage{amsmath,amssymb}
\usepackage[colorlinks=true,allcolors=blue]{hyperref}
\usepackage{dcolumn}
\usepackage{units}
\usepackage{graphicx}
\usepackage{float}
\usepackage{enumitem}
\usepackage{diagbox}
\usepackage{overpic}
\usepackage{wasysym}
\usepackage{comment}
\usepackage{color}
\usepackage{longtable}
\usepackage{lineno}

\allowdisplaybreaks[1]

\newcommand{\be}{\begin{equation}}
\newcommand{\ee}{\end{equation}}

\newcommand{\eref}[1] {Eq.~(\ref{#1})}

\newcommand{\sref}[1] {Sec.~\ref{#1}}
\newcommand{\fref}[1] {Fig.~\ref{#1}}
\newcommand{\aref}[1] {Appendix~\ref{#1}}